\documentclass[lettersize,journal]{IEEEtran}
\IEEEoverridecommandlockouts
\usepackage{cite}
\usepackage{amsmath,amssymb,amsfonts,amsthm}

\usepackage{algorithmic}
\usepackage{graphicx}
\usepackage{textcomp,subcaption}
\usepackage{xcolor,multicol,multirow}
\usepackage{enumerate,textcomp,array,enumitem}
\usepackage{placeins}
\usepackage{threeparttable,url}
\usepackage{dblfloatfix}
\usepackage{flushend}
\usepackage{setspace}
\usepackage{soul}
\usepackage[running,displaymath]{lineno}
\usepackage{float}
\usepackage{placeins}
\usepackage{lineno}

\def\BibTeX{{\rm B\kern-.05em{\sc i\kern-.025em b}\kern-.08em
    T\kern-.1667em\lower.7ex\hbox{E}\kern-.125emX}}

\title{Lunar Power Grid: Network Structure and Spontaneous Synchronization} %

\author{
M.~Walth, \and
A.~Sajadi, \and
M.~Carbone, \and
and B.~M.~Hodge
\thanks{M.~Walth is with the Mathematics Department at Cornell University, Ithaca, NY 14810, USA, and for a part of this research was with the National Renewable Energy Laboratory (NREL), 15013 Denver W Pkwy, Golden, CO 80401, USA, Email: msw283@cornell.edu.}

\thanks{A.~Sajadi is with the Renewable and Sustainable Energy Institute (RASEI) at the University of Colorado Boulder, 4001 Discovery Drive, Boulder, CO 80303, USA, and the National Renewable Energy Laboratory (NREL), 15013 Denver W Pkwy, Golden, CO 80401, USA, Email: Amir.Sajadi@colorado.edu, Amir.Sajadi@nrel.gov.}

\thanks{M.~Carbone is with the National Aeronautics and Space Administration (NASA) Glenn Research Center, 21000 Brookpark Rd, Cleveland, OH 44135, USA, Email: marc.a.carbone@nasa.gov.} 

\thanks{B.~M.~Hodge is with the Department of Electrical, Computer and Energy Engineering, the Renewable and Sustainable Energy Institute (RASEI), and the Department of Applied Mathematics at the University of Colorado Boulder, 425 UCB, Boulder, CO 80309, USA, and the National Renewable Energy Laboratory (NREL), 15013 Denver W Pkwy, Golden, CO 80401, USA, Email: BriMathias.Hodge@colorado.edu, Bri.Mathias.Hodge@nrel.gov.}
}

\begin{document}

\maketitle

\markboth{Preprint Draft, \today}%
{Sajadi \MakeLowercase{\textit{et al.}}: TBD}


\begin{abstract} 

Achieving stable synchronized operation in an alternating current power network is critical to the continuity and reliability of energy delivery. In this paper, we study a dynamic model for synchronization in the proposed power network which the National Aeronautic and Space Agency plans to build on the lunar surface to support continuous human presence on the Moon and a lunar economy. This network is quite remarkable in the sense that it is expected to be the first power network to operate at the unprecedented operating frequency in the $\text{kHz}$ range. The particular structure of this network allows us to derive the necessary and sufficient conditions guaranteeing the existence of a unique locally stable synchronized mode, which will provide a passive control mechanism for the system. Furthermore, we study the bifurcation process leading to the loss of synchronization when the system parameters fall outside of the stable regime. Our results have broader implications for many complex networks with parametric heterogeneity to enhance stability and resilience.
\end{abstract}

\vspace{-.5em}


\vspace{-1em}



\section{Introduction}
Among the myriad dynamical phenomena observed in nature, spontaneous synchronization of complex systems is among the most intriguing, and has received substantial research attention in recent years \cite{ghoshal2011ranking,gao2016universal,pang2018effect,nokkala2016complex,nokkala2018local,perrings1998resilience,kalisch2019deconstructing,mari2015complex,D_rfler_Synch_in_complex_networks,sajadi2022synchronization,motter2013spontaneous,Strogatz_Kuramoto_Review}. Of the systems which display this synchronization phenomena, power grids are of particular interest for several reasons. As high dimensional dynamical systems, power grids have the potential to display extremely complex behaviors, including chaos \cite{Kopell_Chaos_In_Swing}, quasi-periodic behavior \cite{QuasiPeriod_Swing}, and spontaneous synchronization \cite{Dorfler_Kuramoto_Grid}, and as such, power grids are a rich context for discovery of dynamical phenomenon. Therefore, as critical infrastructure for all aspects of modern life, developing insight into their underlying physics is essential to assure their synchronization and steer the system far away from parameter regimes in which undesirable dynamics could arise \cite{Motter_Spontaneous_Synchrony_2013}. Synchronization in an alternating current (AC) power grid refers to the ability of generators to produce power at the same electric frequency, guaranteeing the stable flow of electric power across the network and yielding a uniform frequency at all nodes 
\cite{Bri_and_Amir_2022}. 
Recent recent literature has significantly focused on the synchronization problem, considering the changing nature of electrical generation fleets towards renewable technologies. \cite{Bri_and_Amir_2022,Dorfler_Kuramoto_Grid,Motter_Spontaneous_Synchrony_2013,Monshizadeh_Port_Hamiltonian,Monshizadeh_Port_Hamiltonian_2,Kasis_Passivity_Load_Participation,Trip_Burger_Passivity_Variable_Voltages} Approaches to the problem include modeling the grid as a second-order Kuramoto model \cite{Dorfler_Kuramoto_Grid,D_rfler_Synch_in_complex_networks}, master stability function formalism, \cite{Motter_Spontaneous_Synchrony_2013}, \cite{Motter_Heterogeneity}, parametric sensitivity analysis \cite{Bri_and_Amir_2022},  passivity based arguments from control theory \cite{Kasis_Passivity_Load_Participation, Trip_Burger_Passivity_Variable_Voltages} Port-Hamiltonian systems \cite{Schiffer_Port_Hamiltonian, Monshizadeh_Port_Hamiltonian_2, Monshizadeh_Port_Hamiltonian}. 

We find that the existing body of literature for synchronization in power networks is limited to convergence about $50-60\text{Hz}$. However, the National Aeronautics and Space Administration (NASA) has designs to build a lunar power grid with a ``hub-and-spokes" topology, which they propose operating at a frequency two orders of magnitude higher, in the kilohertz (kHz) frequency range\cite{carbone2023multi, thomas2023lunar} - a regime which has never been studied, tested, or operated in any terrestrial power network. We study this unprecedented power network and aim to identify the constraints bounding the operational parameters for such a system to guarantee stable operation. We study this topology and find particularly tractable analytic conditions for stable synchronization, with very few assumptions on the system parameters. Our central result finds that there is a ``critical coupling" strength between the generators, defined as a simple function of a generator's power injection and damping constant, which must be maintained for stable synchronization to persist. We formulate the necessary and sufficient  condition in an especially concise manner, and formally verify the stability of the synchronized mode using an energy-like Lyapunov function approach. Our results provide a passive control scheme, in the sense that stability is guaranteed without the use of any active feedback monitoring. Additionally, we analyze the bifurcation process that leads to a loss of stable synchronization. We discover that at the bifurcation parameter, the system undergoes an \emph{infinite-period} bifurcation \cite{Strogatz_3rd_ed_2024}. This is an especially dramatic bifurcation, in which a stable fixed point is replaced by a large amplitude oscillation. The implication for the operation of the power network is that outside of the parameter range which guarantees stable synchronization, the network will experience violent, large amplitude oscillations.

We argue that the findings of this paper have broader impacts beyond this particular network, and that our results could be used to control many complex, strongly heterogeneous networks which form a hub-and-spokes topology.




\section{Results}


\subsection{\label{sec:Lunar-Grid-Operation}Lunar Power Grid: Design Structure}

\begin{figure} [ht]
    \centering   
         \includegraphics[width=.47\textwidth]{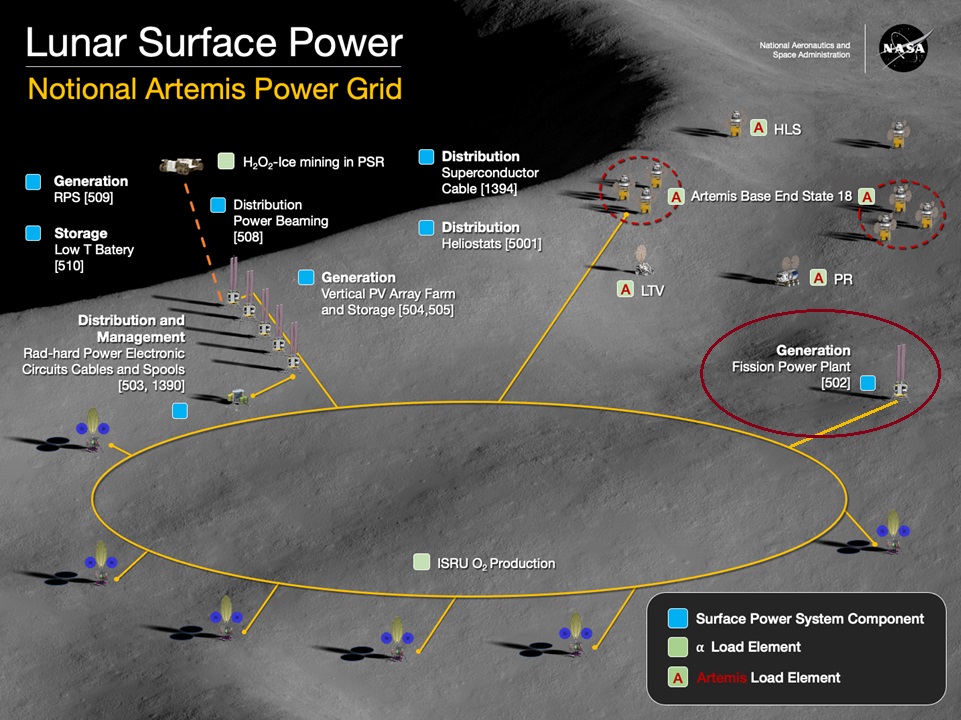}  
         \caption{\textbf{Conceptual depiction of the lunar power grid.}  The design features a large fission nuclear generation unit that is coupled to a number of relatively smaller solid state solar and battery technologies \cite{csank2022power}. The expected loads are the human habitations - whether on the surface or in lunar pits \cite{carrer2024radar} - rover charging for surface exploration and scientific purposes, mining inside lunar craters, and space vehicle landing and take-off sites. The lunar power grid will have comparatively limited generation capacity, due to the cost of deployment and the limitations on availability of resources (because of long lunar nights), requiring source and load power levels to be closely matched \cite{thomas2023lunar}. These features make the lunar grid especially susceptible to instability \cite{carbone2019bifurcation,carbone2023voltage}. Additionally, the sensitive and critical nature of loads, and the need to serve loads in permanently shadowed regions of the Moon, warrant a high degree of robustness to prevent interruptions. 
          } 
  \label{fig:Lunar_grid}   
\end{figure}

The lunar power grid is the first of its kind network whose conceptual layout is pictured in Fig. \ref{fig:Lunar_grid}. It will use AC transmission lines to distribute power over a wide geographical terrain on the surface of an astronomical body that is not Earth in order to support space operations. \cite{oleson2022deployable}. NASA has determined that operating the AC grid in the typical $50-60\text{Hz}$ range would lead to unacceptable levels of power losses, due to inductive coupling with, and dissipation into, the lunar regolith \cite{thomas2023lunar}. For this reason, the proposed lunar grid would operate in the $1-3\text{kHz}$ range, which is nearly unheard of in terrestrial power grids. 
The central node in the formation is expected to be a 40 kW fission nuclear reactor \cite{csank2023nasa,oleson2022deployable} that provides a robust frequency regulation at its point of interconnection and is supported by solar and battery technologies at its adjacent links. The vast majority of equipment and components in the lunar power grid are expected to be solid state generators, in particular, solar based. 
Of the viable solutions to meet the stringent requirements for operating this system, passive control approaches have advantages as a potential operational resilience framework, as the reduced complexity in the controller lends itself to the strict testing and validation requirements of NASA's flight software to ensure the reliability and robustness of the design \cite{timmons2021Core}.
Passive control approaches also reduce the amount of sensors and communications equipment needed to operate a lunar grid, thus decreasing cost and mass for the mission. Most importantly, passive control approaches are shown to be more robust to communications outages, increasing the network's reliability. In this paper, we seek to develop design principles which can be used to passively ensure stable synchronization in the heterogeneous power grid. A more comprehensive description of planned lunar power grid is provided in Supplementary Note 1.

\subsection{Network Topology}

\begin{figure*} [t]
    \centering   
   \begin{subfigure}{0.46\textwidth}
         \centering
         \includegraphics[width=\textwidth] 
         {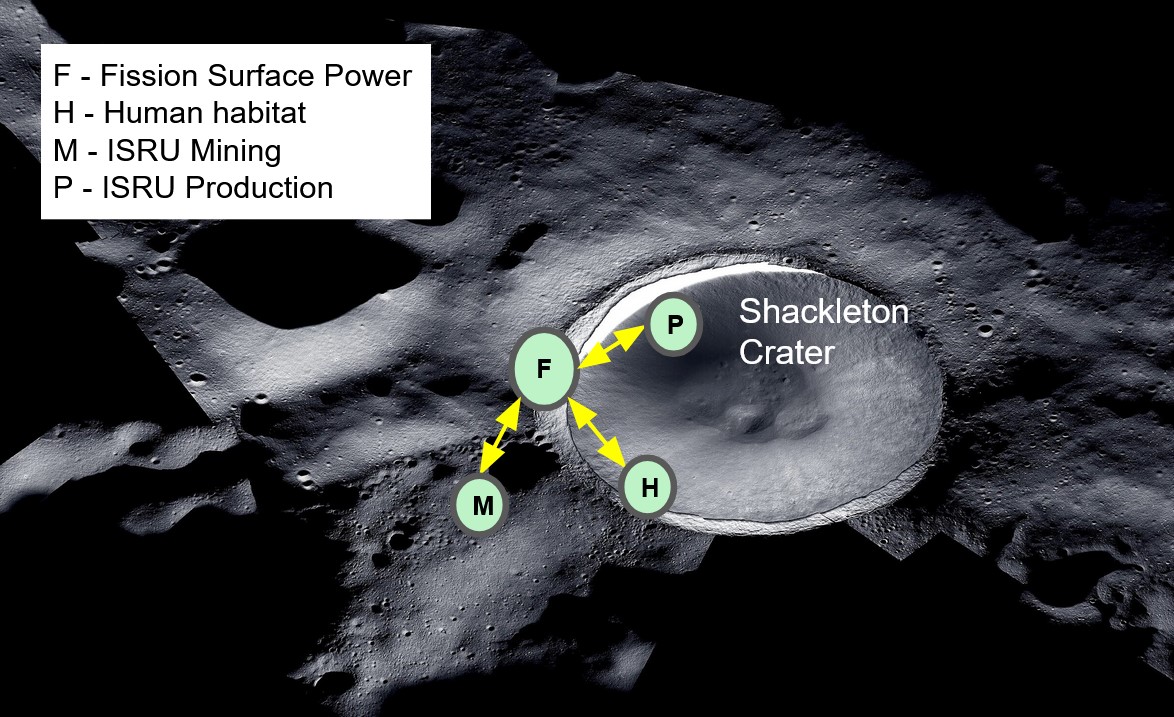}
           \caption{~}   
           \label{fig:crater}
    \end{subfigure} 
     \begin{subfigure}{0.46\textwidth}
         \centering
         \includegraphics[width=\textwidth]     {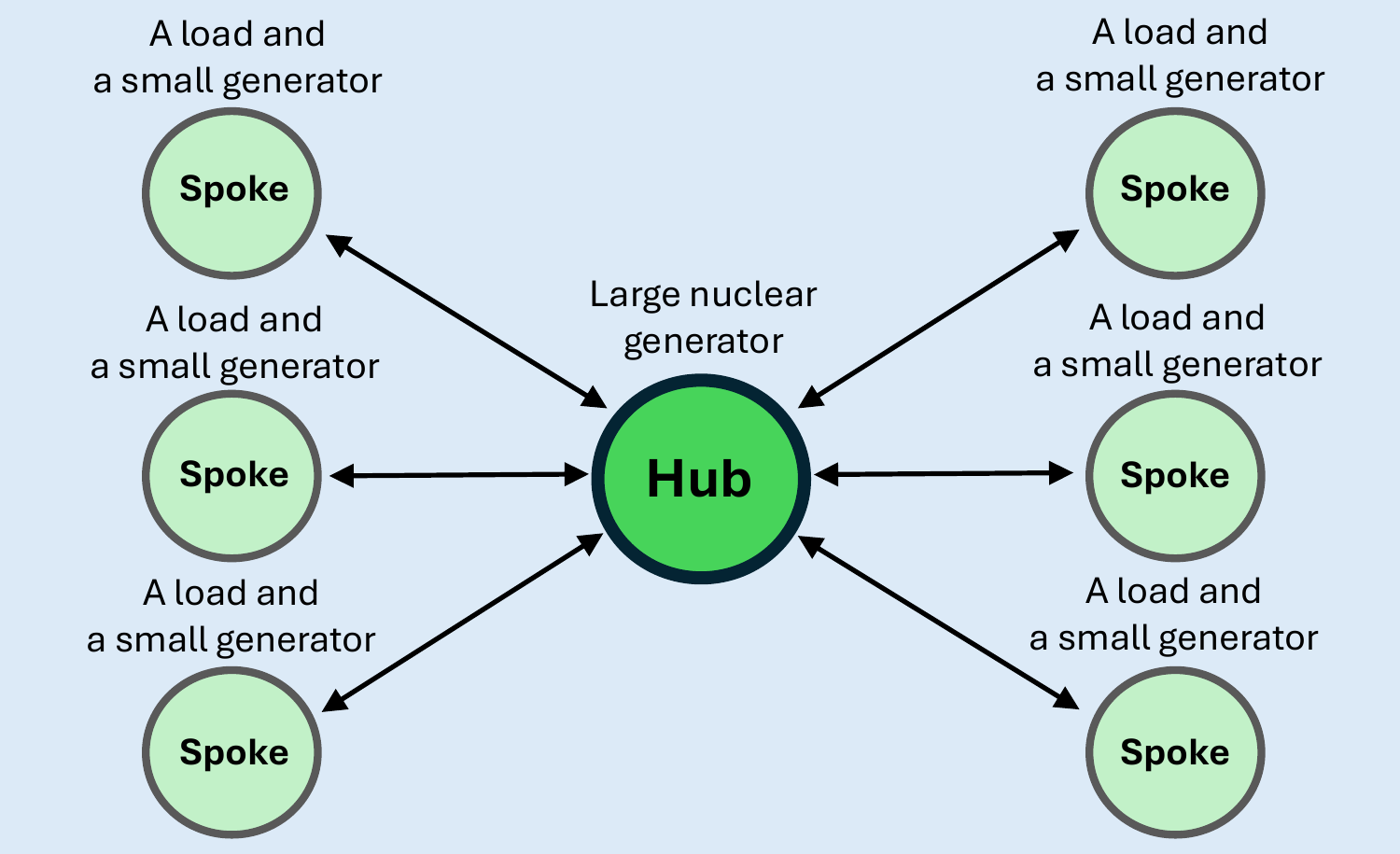}   
         \caption{~}
         \label{fig:graph_toplogy}
    \end{subfigure}    
    \caption{\textbf{Network topology for the lunar power grid.} 
    \textbf{(\ref{fig:crater})} Shackleton Crater that lies at the lunar South Pole - the potential site of the first lunar power grid - and approximate location of generators and loads. This graphic is produced using publicly available lunar images by NASA. The location of components are estimated according to the NASA's trade studies \cite{thomas2023lunar}. 
    \textbf{(\ref{fig:graph_toplogy})} General topology of a Hub-and-spoke network which is NASA has found to be the mass-optimal design for the lunar power network \cite{thomas2023lunar}, and the focus of this study.
A large fission nuclear generation unit in the middle as the hub provides a robust inertial dominance through its centrality. It is coupled with relatively smaller solid state solar and battery technologies and loads as its spokes. Each load will feature its own local small power generation and energy storage. Loads include a human habitat, in situ resource utilization (ISRU) mining operations, and an ISRU production facility which will process mined materials into usable materials. Note that although this proposed design is suggested by the trade studies \cite{thomas2023lunar, thomas2022establishing}, this design does not imply or reflect an official endorsement from NASA. Image Credit: LROC (Lunar Reconnaissance Orbiter) and ShadowCam teams with images provided by NASA/KARI/ASU \cite{nasa_image}.
}
  \label{fig:Hub-Spokes}   
\end{figure*}

NASA's trade studies investigated a variety of network topologies \cite{thomas2023lunar, thomas2023modular} and found that the network topology which optimizes the amount of power a grid can support for a given amount of mass is given by a radial network \cite{thomas2023lunar, thomas2023modular}, as shown in \ref{fig:Hub-Spokes}(a). Proposals therefore suggest that the initial stage of deployment uses a radial network, akin to a hub-spoke topology, with expected expansion into a larger formation in modular fashion to serve multiple functional zones \cite{thomas2022establishing, thomas2023modular}. We therefore focus our study in this paper on a hub-and-spokes topology, as depicted in Fig. \ref{fig:Hub-Spokes}(b).

\subsection{\label{sec_sup:Model-and-Assumptions}Dynamic Model}

The dynamics of coupled AC generators are governed by the so-called swing equations \cite{Motter_Comparative_Analysis_of_Power_Grid}. In the case of $n$ coupled generators, the swing equations are order $2n$. We let generator $1$ represent the large synchronous fission reactor, and assume that generators $2,\ldots,n$ are solid state generators, which have very low inertia \cite{Wallace_Bri_and_Amir_Review}. We assume that the solid state generators are strongly coupled to the fission reactor, and negligibly coupled to one another. The proposed grid will feature one load and one small generator at each ``spoke," as described in \cite{thomas2023lunar} and depicted in Fig. \ref{fig:Hub-Spokes}(b).

After performing some simplifications, laid out in Supplementary Notes 2 through 4, the governing equations can be written as:

\begin{equation}
\begin{aligned}\dot{\delta}_{i} & =\Delta\omega_{i}-\Delta\omega_{n}\text{, } \hspace{10pt} i=1,\ldots,n\\
\Delta\dot{\omega}_{1} & =\frac{\omega_R}{2H_1}\left (-D_{1}\Delta\omega_{1}+A_{1}-\sum_{j=2}^{n} K_{1j}\sin(\delta_{1} - \delta_{j})\right )\\
\Delta\dot{\omega}_{i} & =\frac{\omega_R }{2H_i}\left(-D_{i}\Delta\omega_{i}+A_{i}- K_{i1}\sin(\delta_{i} - \delta_{1})\right )\text{, }  \hspace{2pt} i\neq 1.
\end{aligned}
\label{eq:Swing_Equation-simplified}
\end{equation}

Here, $\delta_{i}$ and $\omega_{i}$ represent the phase angle and frequency of the power output of generator $i$, respectively, and $\omega_R$ represents the system's reference frequency, which we take to be $1\text{kHz}$ in the lunar grid. The power angle of machine $i$ in this model is given by $\delta_i +\omega_R t$. We use the notation $\Delta \omega_i := \omega_i -\omega_R$ to represent the frequency deviation of machine $i$ from the reference frequency. A stable synchronized mode is defined to be one in which \cite{sajadi2022synchronization}
\begin{equation}
    \dot{\delta}_1 = \dot{\delta}_2=\cdots = \dot{\delta}_n
\end{equation}
To develop the model shown in Equations (\ref{eq:Swing_Equation-simplified}), we have performed a coordinate transformation $\delta_i \to \delta_i-\delta_n$ so that a synchronized mode corresponds to a fixed point of these equations, which introduces the additional term $\Delta \omega_n$ in the $\dot{\delta_i}$ equations; details can be found in Supplementary Note 4.
The parameter $H_i$ represents the inertial constant of generator $i$. The damping coefficient of generator $i$ is represented by $D_{i}$, and the term $A_{i}$ represents the net power export of $i$th generator to the network. The parameter $A_i$ can be either positive or negative, corresponding to whether node $i$ is generating more power than its load is consuming or vice versa. We treat $A_i$ as constant for the sake of this analysis. The term
$K_{ij}$ corresponds to the overall coupling strength between generators $i$ and $j$. The value of $K_{ij}$ is determined by factors such as the length of the line connecting two generators, the materials used in the construction of the line, whether the line is above or below ground, etc. 

We note that our subsequent analysis applies to any number of solid state generators, making the results relevant to the long term development of the lunar grid, as more generators and loads are added in a modular fashion \cite{thomas2023modular}. More details about the interpretation and derivation of this dynamic model can be found in Supplementary Note 2.
%



We are able to derive that if the generators synchronize, then the synchronization frequency will be
  \begin{equation} 
\label{eq:sync_frequency_2}\Delta\omega_{sync}=\frac{\sum_{j=1}^{n}A_{j}}{\sum_{j=1}^{n}D_{j}},\end{equation}
or equivalently 
\begin{equation}
    \omega_{sync} = \omega_R + \frac{\sum_{j=1}^{n}A_{j}}{\sum_{j=1}^{n}D_{j}}.
\end{equation}

A formal proof is provided in Supplementary Note 5. We note that the term \eqref{eq:sync_frequency_2} agrees with the synchronization condition  found in \cite{D_rfler_Synch_in_complex_networks} in the context of smart grid networks.
Equation (\ref{eq:sync_frequency_2}) confirms our intuition about frequency in AC power networks: namely, the deviation from the reference frequency is proportional to the difference between the power produced by generators and the power consumed by the loads. Excess power generation will tend to increase the frequency, while excess power consumption will tend to drive the frequency downward.
Subsequently, we find the phases
$\delta_{i}$ associated with a fixed point to be: 

\begin{equation}
\delta_{1} = (-1)^k \sin^{-1}\left( \mu_1\right) + k\pi \quad \text{for }k=0,1
\label{eq:(delta1)}
\end{equation}
\begin{equation} \label{eq:Fixed_Point_delta_i}
\begin{split}
\delta_i &= \delta_1 + (-1)^{j+1} \sin^{-1}(\mu_i) -j \pi  \quad \text{for }j=0,1
\end{split}
\end{equation}

where 
\begin{equation}
\mu_{i}:=\frac{D_{i}\Delta \omega_{sync}-A_{i}}{K_{i1}} \text{ for } i>1
\end{equation}

and 
\begin{equation}
    \mu_1 =\frac{D_{n}\Delta \omega_{sync}-A_{n}}{K_{n1}}.
\end{equation}

Together, equations \eqref{eq:sync_frequency_2}, \eqref{eq:(delta1)}, \eqref{eq:Fixed_Point_delta_i} determine the locations of all of the fixed points of the swing equations (\ref{eq:Swing_Equation-simplified}). Note that this implies the existence of $2^{2n-2}$ total fixed points.

\begin{figure*} [t]
    \begin{subfigure}{0.245\textwidth}
         \centering
         \includegraphics[width=\textwidth]
        {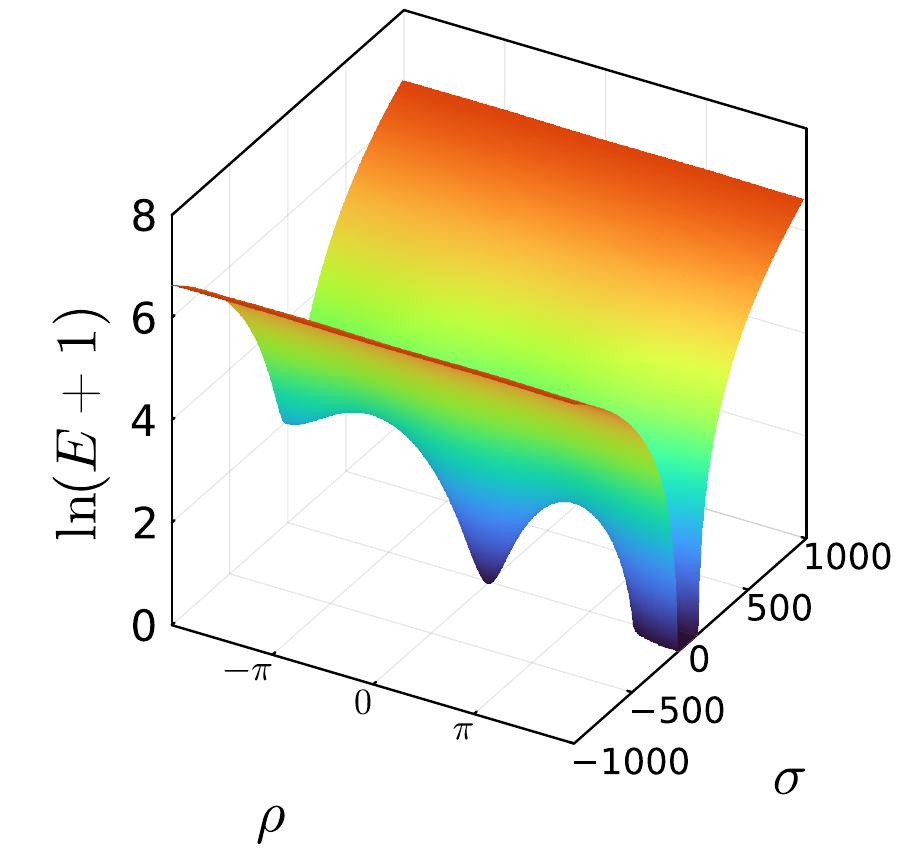}  
    \end{subfigure} 
%
%
    \begin{subfigure}{0.245\textwidth}
         \centering
         \includegraphics[width=\textwidth]{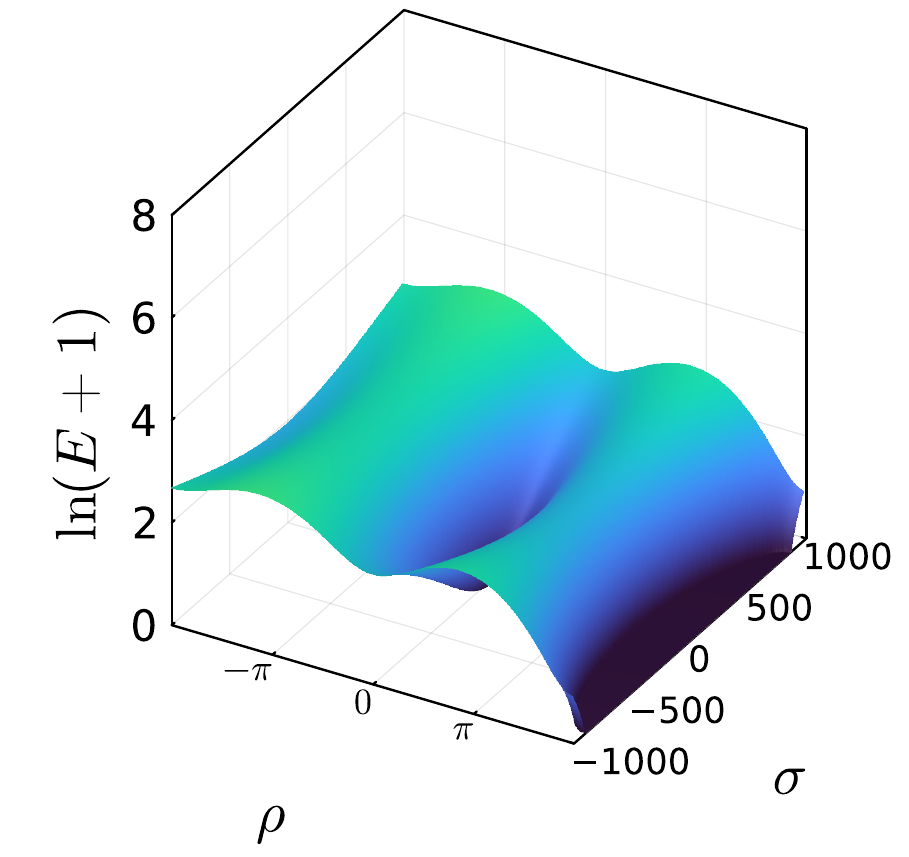}  
    \end{subfigure}
 %
%
    \begin{subfigure}{0.245\textwidth}
         \centering
         \includegraphics[width=\textwidth]{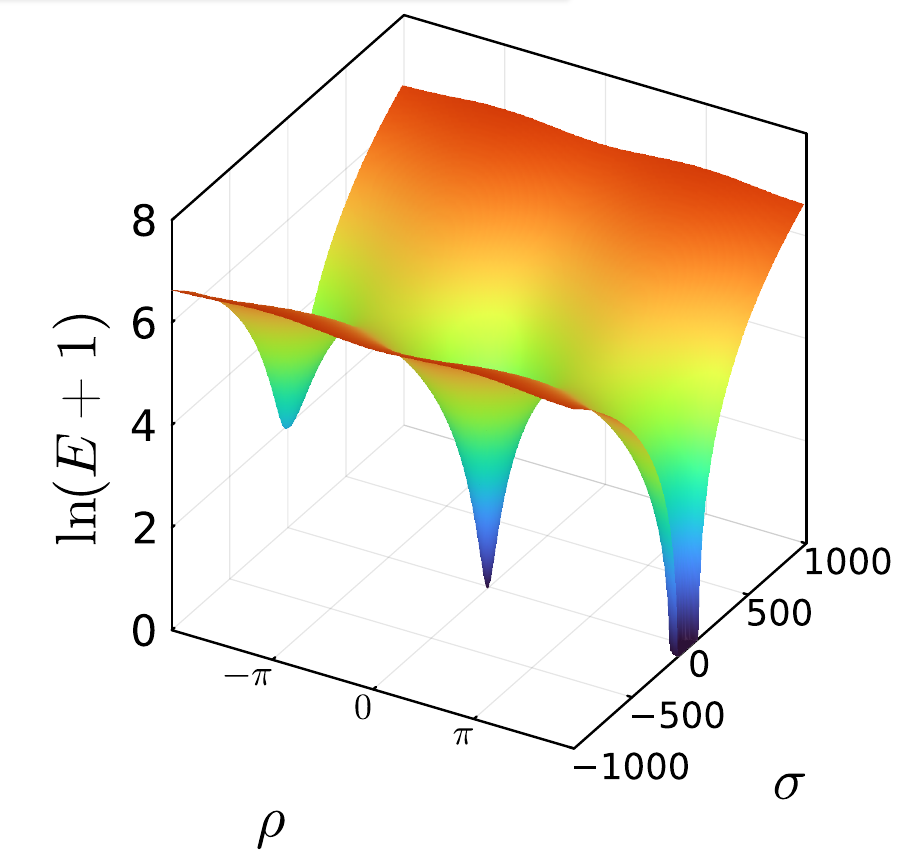}  
    \end{subfigure}%
%
%
    \begin{subfigure}{0.245\textwidth}
         \centering
         \includegraphics[width=\textwidth]{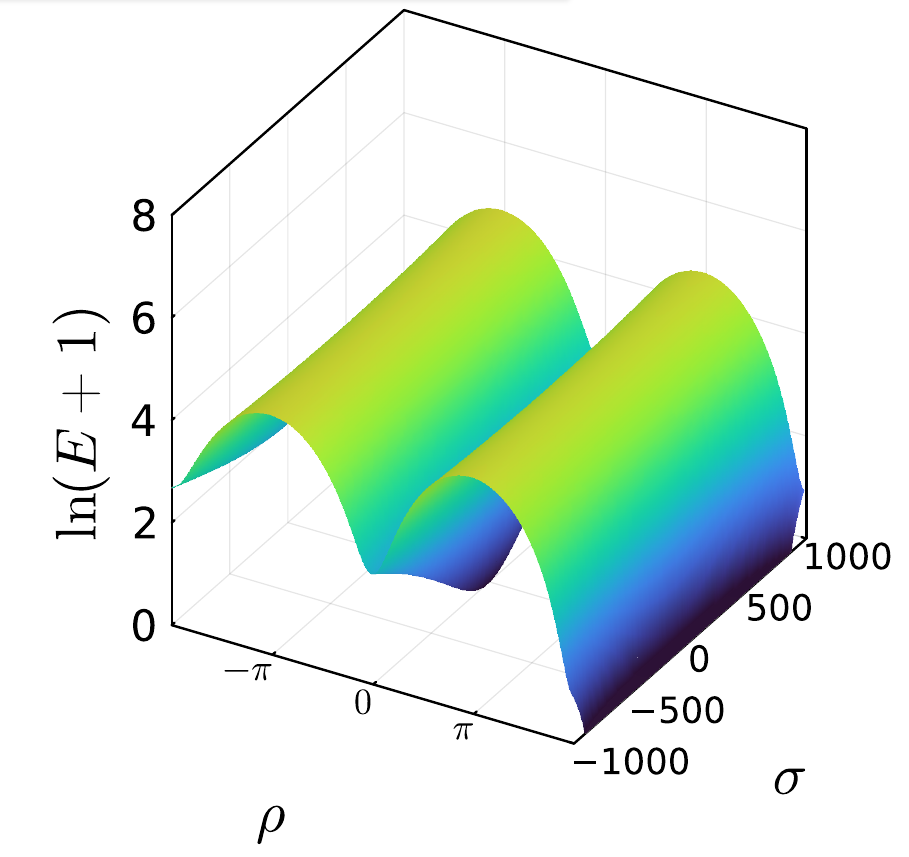}  
    \end{subfigure}
    \begin{subfigure}{0.245\textwidth}
         \centering
         \includegraphics[width=\textwidth]        {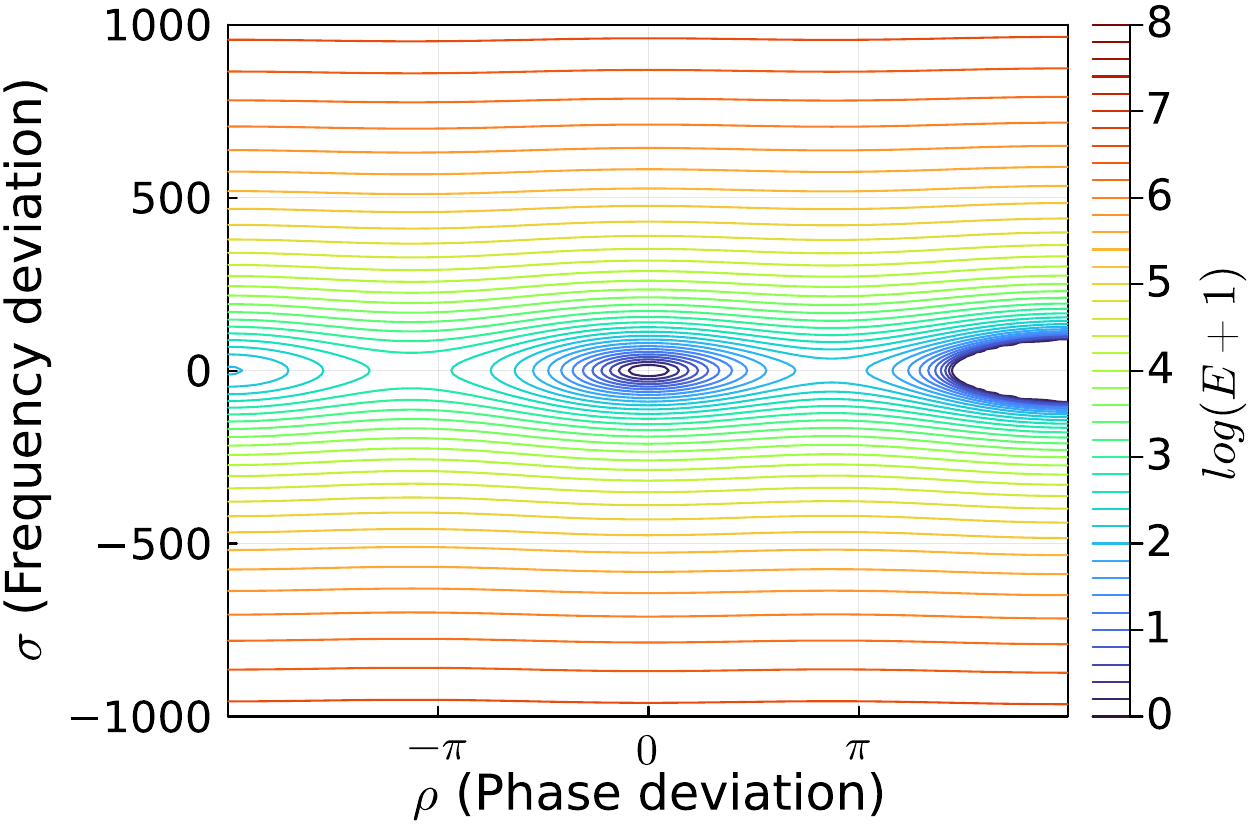}  
           \caption{~}   
           \label{fig:Completely_Homogeneous_Contours}
    \end{subfigure} 
%
     \begin{subfigure}{0.245\textwidth}
         \centering
         \includegraphics[width=\textwidth]{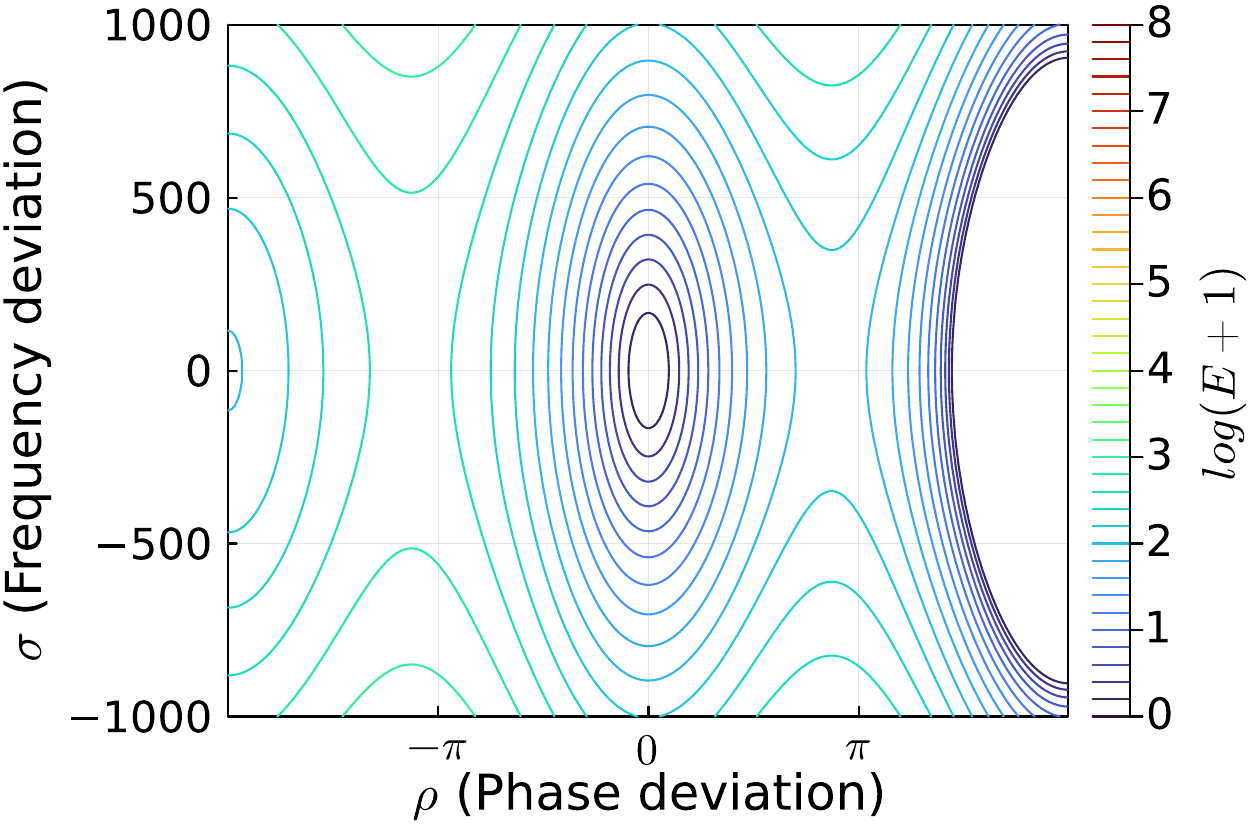} 
         \caption{~}
         \label{fig:H_hetero_Contours}
    \end{subfigure}     
%
     \begin{subfigure}{0.245\textwidth}
         \centering
         \includegraphics[width=\textwidth]{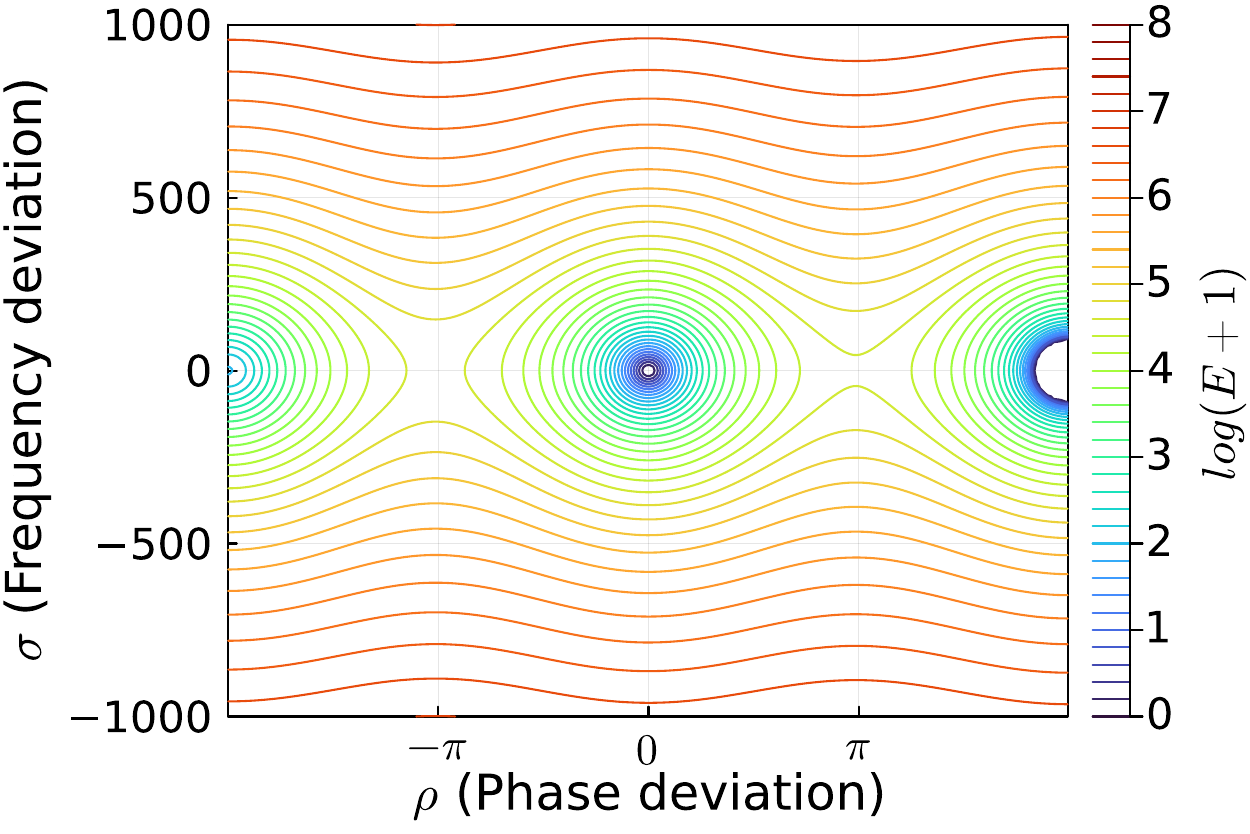} 
         \caption{~}
         \label{fig:Homogeneous_Large_Coupling_Contours}
    \end{subfigure}      
%
     \begin{subfigure}{0.245\textwidth}
         \centering
         \includegraphics[width=\textwidth]{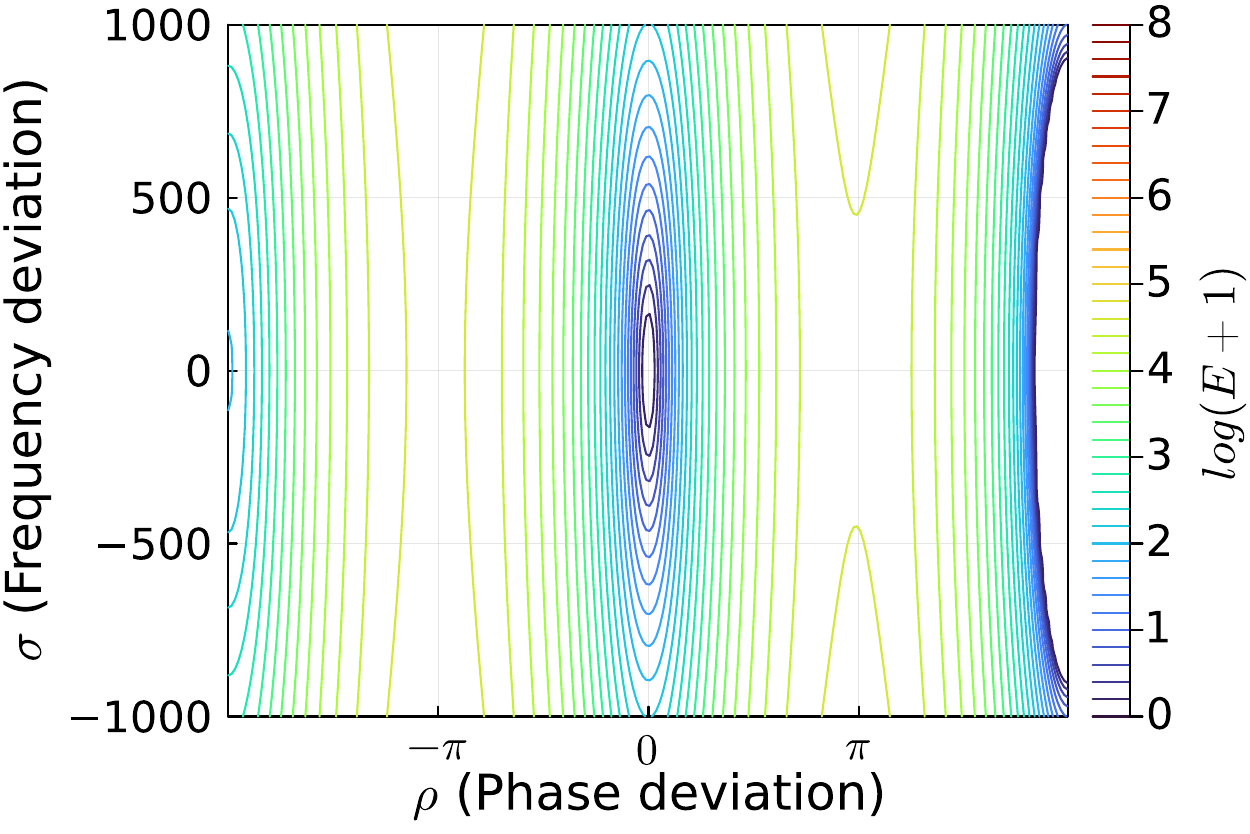} 
         \caption{~}
         \label{fig:H_hetero_Large_Coupling_Contours}
    \end{subfigure}   
    \caption{\textbf{The log of the energy function $E$ is depicted on the vertical axis.} The synchronized state is at $(0,0)$ on the horizontal plane. $\rho$ quantifies deviation from the synchronized state in phase $\delta_2$, and $\sigma$ quantifies deviation in frequency $\Delta \omega_2$. Note that in all cases, the synchronized fixed point is at the bottom of an energy well, indicating that it is stable.
    \textbf{(\ref{fig:Completely_Homogeneous_Contours})} Homogeneous Inertia - Base Coupling. This base case shows the energy function when the generators are identical, and the coupling constants $K_{ij}$ are at a constant baseline value. Note the stable fixed point at $(0,0)$ and the unstable fixed points near $\rho = \pm \pi$.
    \textbf{(\ref{fig:H_hetero_Contours})} Heterogeneous Inertia - Base Coupling. This case illustrates a heterogeneous grid - generator $1$ is taken to be a large, high inertia generator, while the remaining generators are taken to be small solid state generators with very low inertia, $100\times$ smaller than baseline. Note that the lower inertia decreases the overall energy, but has no effect on the stability of $(0,0)$.
    \textbf{(\ref{fig:Homogeneous_Large_Coupling_Contours})} Homogeneous Inertia - Large Coupling - The grid is again taken to be homogeneous, but with coupling $K_{ij}$ increased by a factor of $10$ relative to baseline. Increased coupling increases the overall energy. Notice that the unstable fixed points have moved to almost exactly $\rho = \pm \pi$. For this reason, large coupling increases the radius of the basin of attraction of the stable fixed point.
    \textbf{(\ref{fig:H_hetero_Large_Coupling_Contours})} Heterogeneous Inertia - Large Coupling - Again, a heterogeneous grid, similar to figure \textbf{(b)}, but with coupling increased by a factor of $10$ over baseline. Compared to figure \textbf{(c)}, the lower inertia decreases the overall energy, but the synchronized mode remains stable.
}
     \label{fig:energy_functions} 
\end{figure*}
From equations (\ref{eq:(delta1)}) and (\ref{eq:Fixed_Point_delta_i}), we see that a necessary condition for the existence of synchronized modes is $|\mu_i| \leq 1$ for each $i>1$. Further discussion of this result can be found in Supplementary Note 5.
Thus far, we have found a description of the synchronized fixed points and determined a necessary condition for such fixed points to exist. In the next section, we discuss the stability of these fixed points.


\subsection{Stability of Synchronized Modes}\label{sec:Stability_Of_Synchronized_Modes}

As our main result, we prove that system (\ref{eq:Swing_Equation-simplified}) has a unique stable synchronized fixed point if and only if
\begin{equation} \label{eq:unique_solution_precise_main}
\left |D_i \Delta \omega_{sync}-A_i \right |<K_{1i} ~~~~~
\text{for } i=2,\ldots,n.
\end{equation}
When this condition is met, for all initial conditions other than the unstable fixed points, the system will tend to this stable synchronized fixed point.
This result can be interpreted as saying that so long as the grid has sufficiently robust coupling, the system is guaranteed to tend towards frequency synchronization. The size of the required coupling is determined by the system deviation from its reference frequency, $\Delta \omega_{sync}$, the damping of each small generator $D_i,$ and the net power output/consumption at node $i$. Another interpretation of inequality (\ref{eq:unique_solution_precise_main}) is that it gives an effective bound on how large of an over/under frequency event can occur within the network before synchronized operation is threatened. A complete proof of this result is laid out in the Supplementary Notes 5 and 7, and is summarized here.

To demonstrate the uniqueness of the stable synchronized mode, we begin by linearizing the model
(\ref{eq:Swing_Equation-simplified}) to study the eigenvalues of the system at a fixed point $(\boldsymbol{\delta},\boldsymbol{\Delta \omega})$.
We derive the system Jacobian, $J$, which can be written in block matrix form as: 
\begin{equation*}
J(\boldsymbol{\delta},\boldsymbol{\Delta \omega})=\left [\begin{matrix}
-\alpha_1 & \boldsymbol{0}^T  & 1\\
\boldsymbol{\alpha}  & \Lambda & \boldsymbol{0}\\
\gamma & \boldsymbol{\beta}^T & \lambda_n
\end{matrix}
\right ]
\end{equation*} where $\boldsymbol{0}$ is the $0$ vector of size $n-2$, $\boldsymbol{\alpha}$ and $\boldsymbol{\beta}$ are vectors of length $n-2$, $\Lambda$ is an $(n-2)\times (n-2)$ diagonal matrix, and $\gamma$ and $\lambda_n$ are scalars. 
%
 %
We carry out a complete linear stability analysis in  Supplementary Notes 6, and we find that for sufficiently large coupling, Equations (\ref{eq:Swing_Equation-simplified}) have exactly one stable fixed point, which we denote $(\delta^s,\Delta \omega_{sync}).$

Next, to sharpen the stability result, we derive an energy-like Lyapunov function, given by:
\begin{equation}
    \begin{split} \label{eq:liapunov_total_energy}
E(\delta,\Delta \omega)&= \frac{1}{2} \sum_{j=1}^{n} m_i \cdot \Delta \omega_i ^2 - \sum_{j=1}^{n}  A_i \cdot (\delta_{i} - \delta^{s}_i) \\
&\cdots - \sum_{j=2}^n K_{ij} \cdot [\cos(\delta_1 - \delta_j)- \cos(\delta_1^s-\delta_j^s)].
    \end{split}
\end{equation}

\begin{figure*}[t]

    \begin{subfigure}{0.24\textwidth}
         \centering
         \includegraphics[width=\textwidth]        {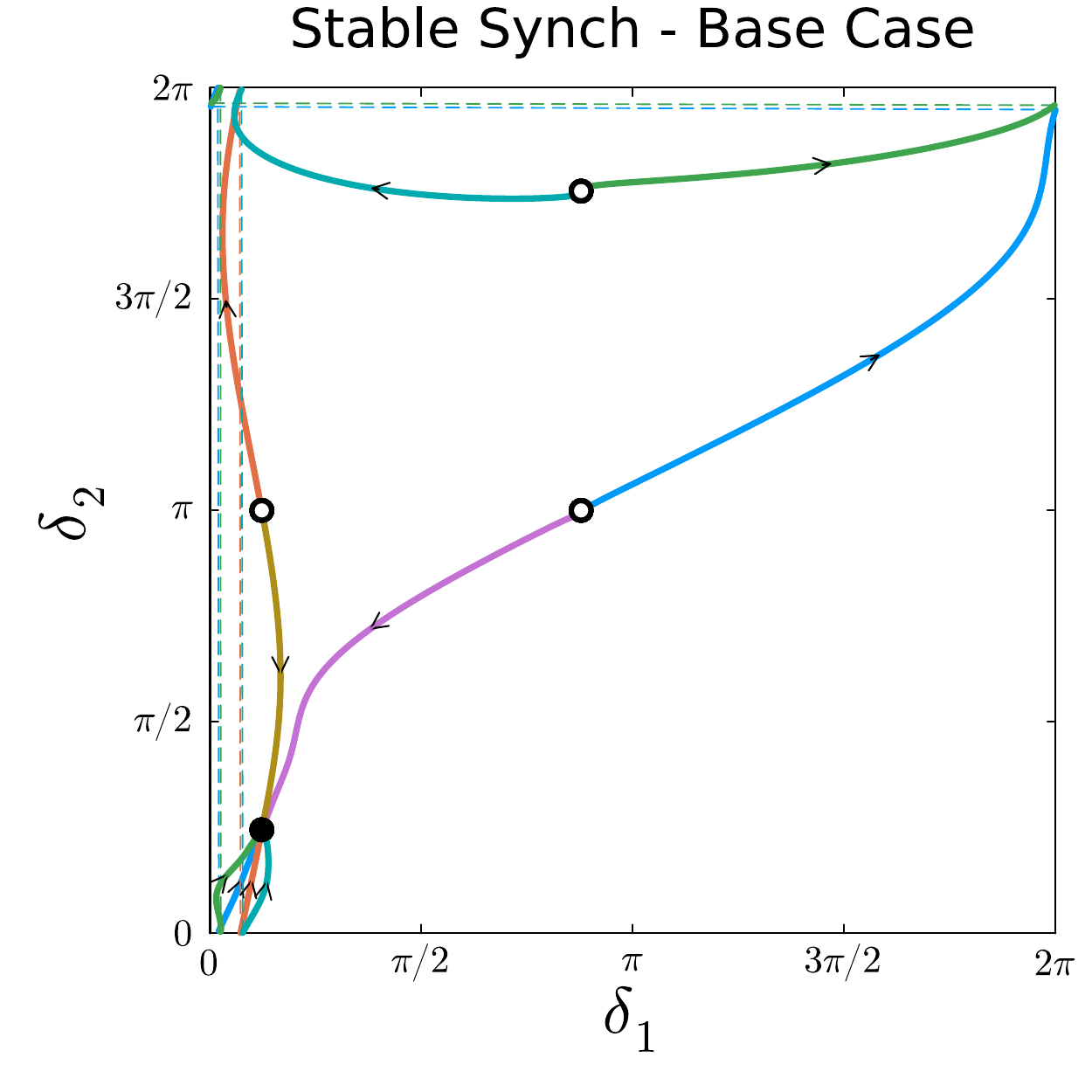}  
           \caption{~}   
           \label{fig:Phase_Plot_Base}
    \end{subfigure} 
     \begin{subfigure}{0.24\textwidth}
         \centering
         \includegraphics[width=\textwidth]{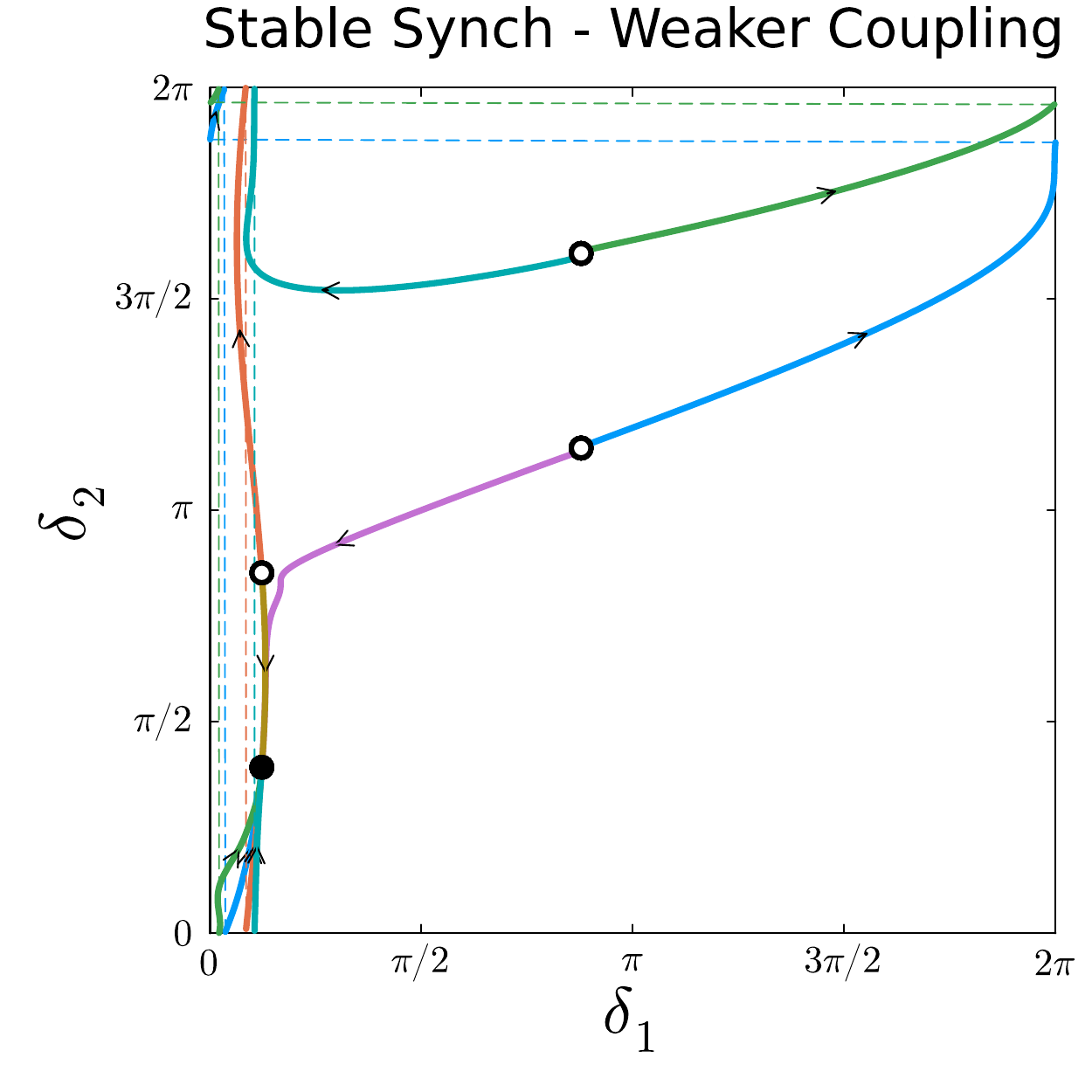} 
         \caption{~}
         \label{fig:Phase_Plot_Weaker}
    \end{subfigure}     
     \begin{subfigure}{0.24\textwidth}
         \centering
         \includegraphics[width=\textwidth]{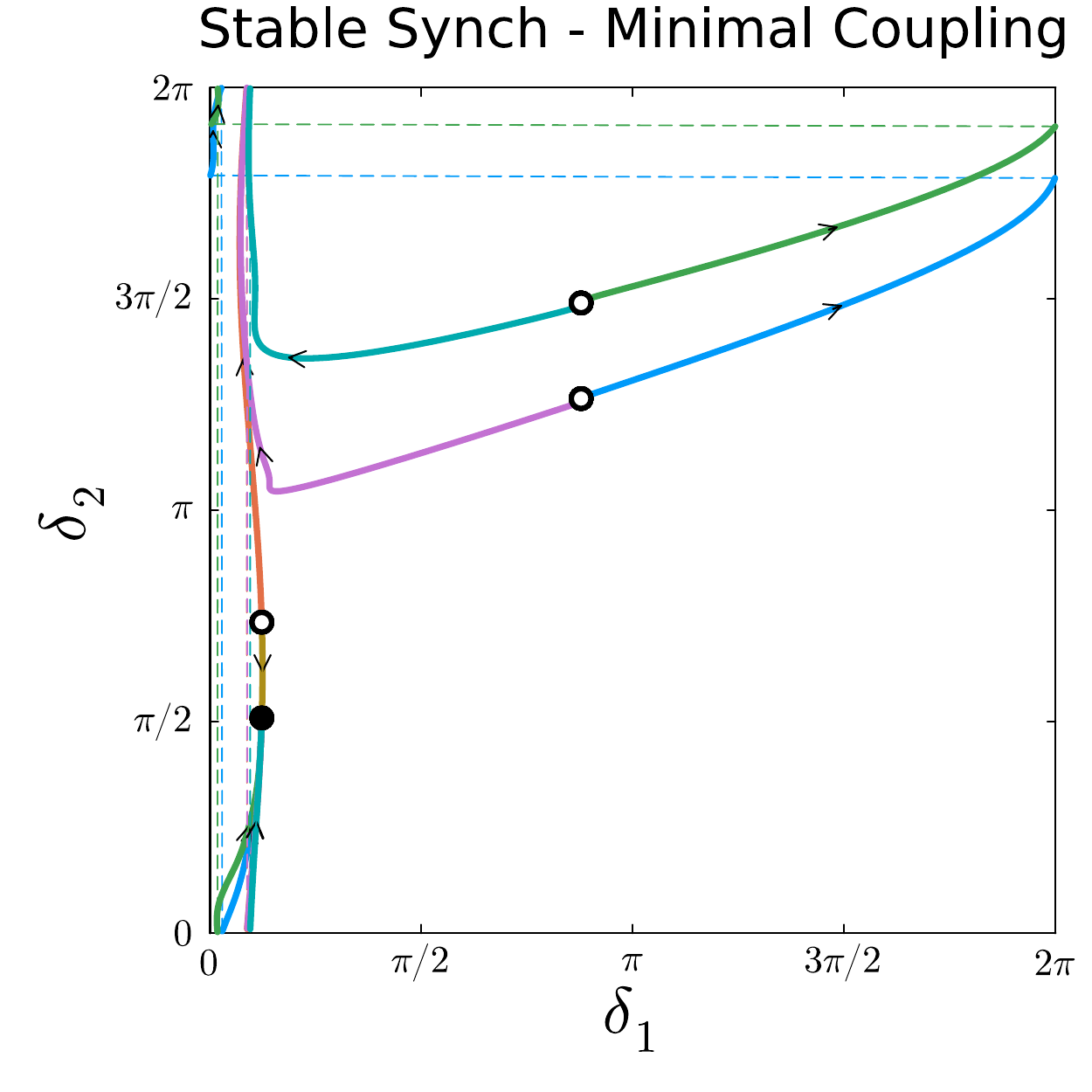} 
         \caption{~}
         \label{fig:Phase_Plot_Weakest}
    \end{subfigure}      
     \begin{subfigure}{0.24\textwidth}
         \centering
         \includegraphics[width=\textwidth]{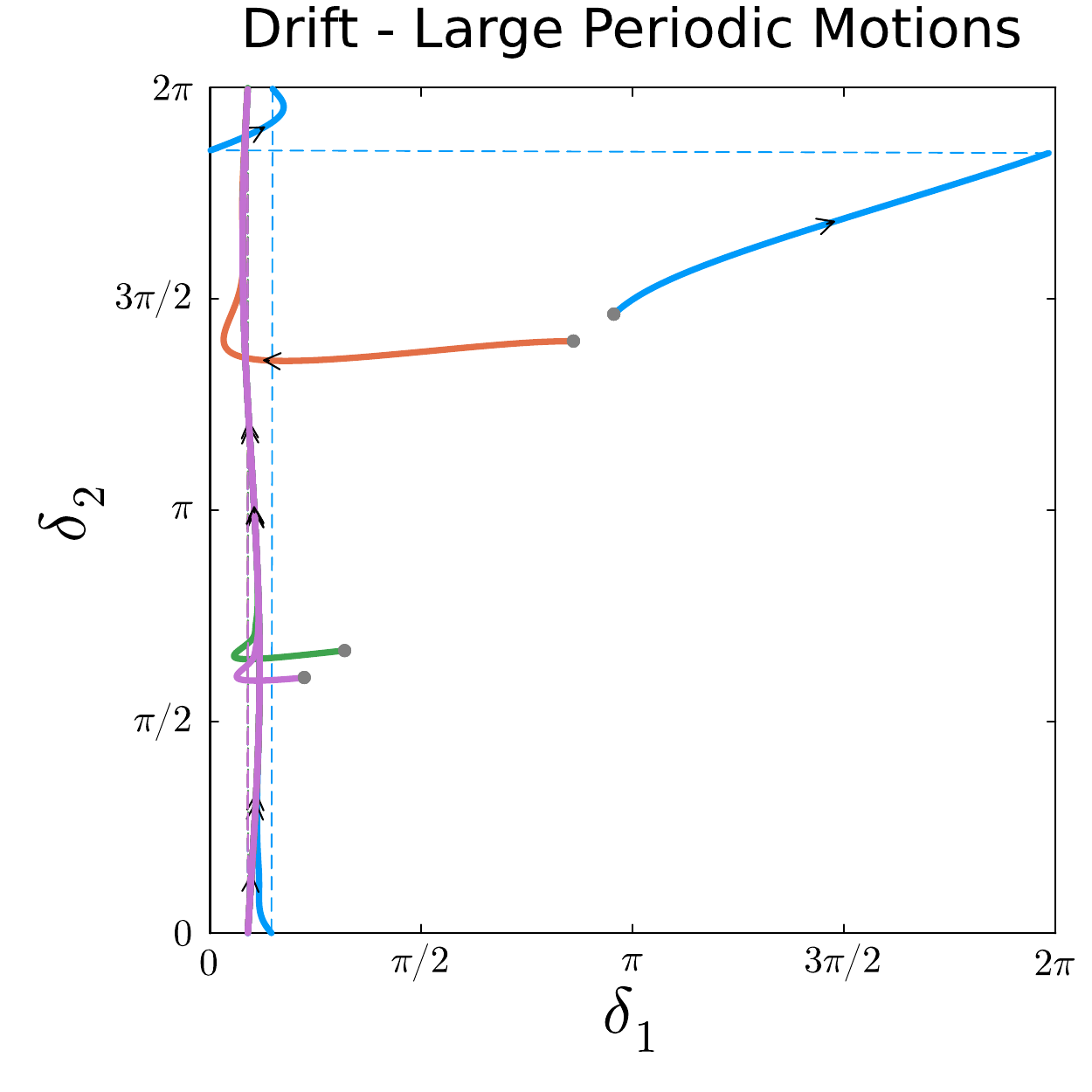} 
         \caption{~}
         \label{fig:Phase_Plot_Drift}
    \end{subfigure}   
    \caption{\textbf{Depiction of the infinite-period bifurcation that destroys the stable synchronized fixed point.}  The system is $2\pi$ periodic - if the plots above were extended to show the range $[-2\pi,2\pi],$ one would see the same patterns repeated every $2\pi$.
    Trajectories are therefore plotted on a torus - the top of each phase plane at $2\pi$ is identified with (or "glued to") the bottom at 0, and the left side is identified with the right. When trajectories wrap around torus, it is denoted with the thin dotted line. The plots were created by simulating a 6 generator system for various parameter values. 
    Hollow circles depict unstable fixed points, and filled circles depict the stable fixed point. In the rightmost graphic, there are no fixed points - instead initial conditions are marked with a small grey dot. Differently colored lines correspond to different initial conditions.
    \textbf{(\ref{fig:Phase_Plot_Base})} In the base case, all trajectories lead to stable fixed point in the bottom left of the phase plane. There are 4 fix points pictured, only one of which is stable.
    \textbf{(\ref{fig:Phase_Plot_Weaker})} Coupling $K_{ij}$ is decreased. Synchronized fixed point is still stable. Notice that fixed points move closer to each other.
    \textbf{(\ref{fig:Phase_Plot_Weakest})}  Coupling $K_{ij}$ is decreased still further. Synchronized state is still stable. Note fixed points are extremely close.
    \textbf{(\ref{fig:Phase_Plot_Drift})} Coupling decrease below critical value. Fixed points have pairwise annihilated in saddle-node bifurcations. Trajectories now wind around the torus in large periodic motions, shown as the vertical curve in the left part of the plane.
    }
     \label{fig:Bifurcation} 
\end{figure*}

We present the derivation of this function in the Supplementary Note 8, where we also prove that $E(\delta,\Delta \omega)$ is in fact a Lyapunov function. The existence of an energy-like function shows that the system will tend to behave similarly to a dissipative mechanical system, in the sense that energy decreases along trajectories, eventually reaching a minimum at a fixed point.
This implies that the fixed point $(\delta^s,\Delta \omega_{sync})$ is locally asymptotically stable whenever it exists, which is precisely when the condition of Inequality (\ref{eq:unique_solution_precise_main}) is satisfied.

Figure \ref{fig:energy_functions} illustrates cross sections of the energy function in several key cases. Because $E$ depends on $2n$ variables, we can only visualize a $2$ dimensional slice of the function. The main feature to notice is that across all subfigures, the synchronized fixed point sits at an energy minimum. 
The plots were generated by plotting $E\big ( (\delta^s ,\Delta \omega_{sync}) +  (\rho \boldsymbol{e_1},\sigma \boldsymbol{e_2}) \big )$, where $\rho$ and $\sigma$ are scalars, $\boldsymbol{e_1}$ is a vector of all zeros except in the position of $\delta_2,$ and $\boldsymbol{e_2}$ is a vector of all zeros except in position $\Delta \omega_2$. Thus, the figure depicts how the energy function changes as generator $2$ deviates from the synchronized fixed point. The variable $\rho$ quantifies the deviation in phase and $\sigma$ quantifies the deviation in frequency. We display the natural log of $E$ because the energy varies over many orders of magnitude.

In this section, we found a precise condition on the parameters of the lunar grid to guarantee stability of synchronized operation. In the following section, we describe the process by which synchronization fails when condition \eqref{eq:unique_solution_precise_main} is not met.


\subsection{\label{sec:Bifurcation} The Bifurcation Process}

So far, we have seen that the lunar grid will synchronize precisely when Inequality \eqref{eq:unique_solution_precise_main} is satisfied. The condition suggests that there is some kind of bifurcation which occurs when $|D_i \Delta \omega_{sync}-A_i|=K_{1i}$ for some $i$. In this section we explore this bifurcation.

\begin{figure*} [!b]

\begin{center}
    \begin{subfigure}{0.16\textwidth}
         \centering
         \includegraphics[width=\textwidth]{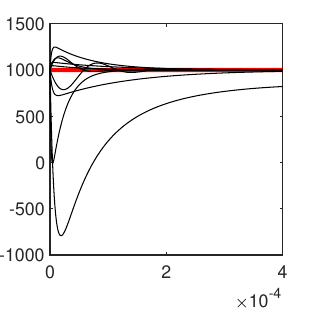}   
    \end{subfigure}
    \hfill
         \begin{subfigure}{0.16\textwidth}
         \centering
         \includegraphics[width=\textwidth]{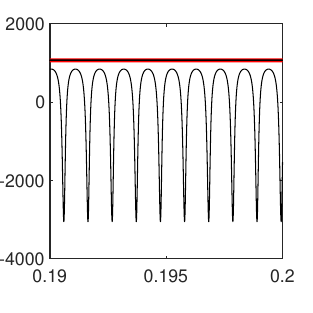}   
         \end{subfigure}
    \hfill
    \begin{subfigure}{0.16\textwidth}
         \centering
         \includegraphics[width=\textwidth]{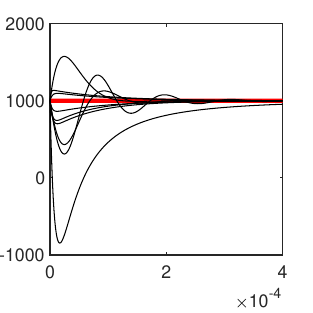}   
    \end{subfigure}
   \begin{subfigure}{0.16\textwidth}
         \centering
         \includegraphics[width=\textwidth]{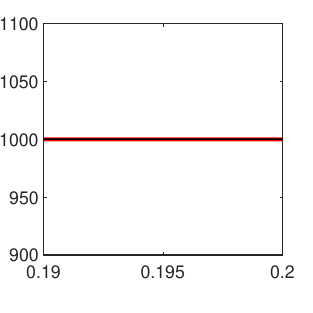}   
    \end{subfigure}
    \hfill
    \begin{subfigure}{0.16\textwidth}
         \centering
         \includegraphics[width=\textwidth]{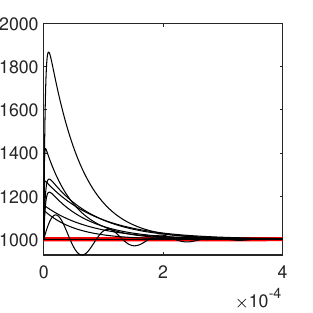}   
    \end{subfigure}
    \hfill
    \begin{subfigure}{0.16\textwidth}
         \centering
         \includegraphics[width=\textwidth]{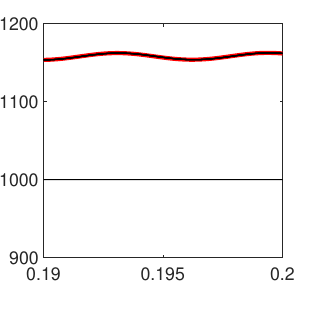}   
    \end{subfigure}
    \begin{subfigure}{0.325\textwidth}
    
         \centering
         \includegraphics[width=\textwidth]{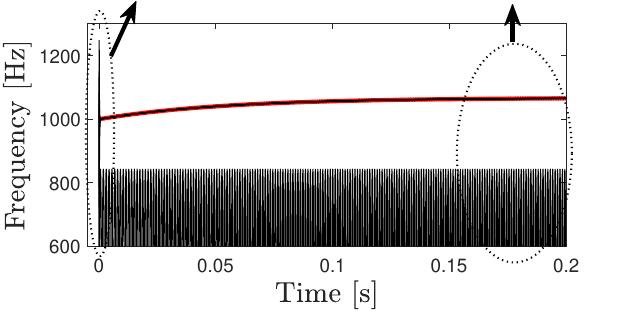}   
    \end{subfigure}
    \hfill
    \begin{subfigure}{0.325\textwidth}
         \centering
         \includegraphics[width=\textwidth]{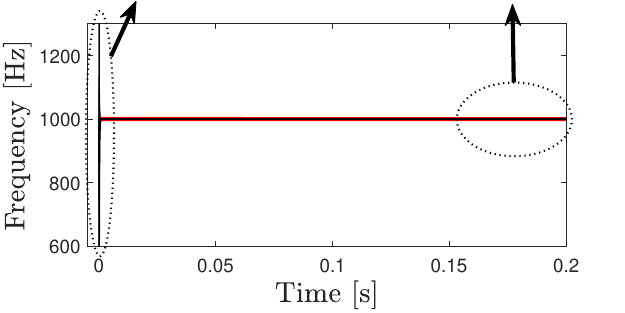}   
    \end{subfigure}
    \hfill
    \begin{subfigure}{0.325\textwidth}
         \centering
         \includegraphics[width=\textwidth]{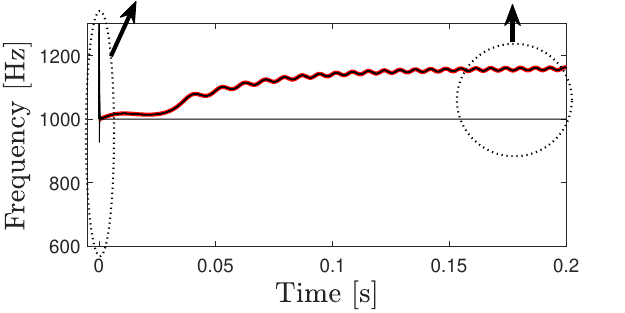}   
    \end{subfigure}
    \hfill
      \begin{subfigure}{\textwidth}
         \centering         \includegraphics[width=.75\textwidth]{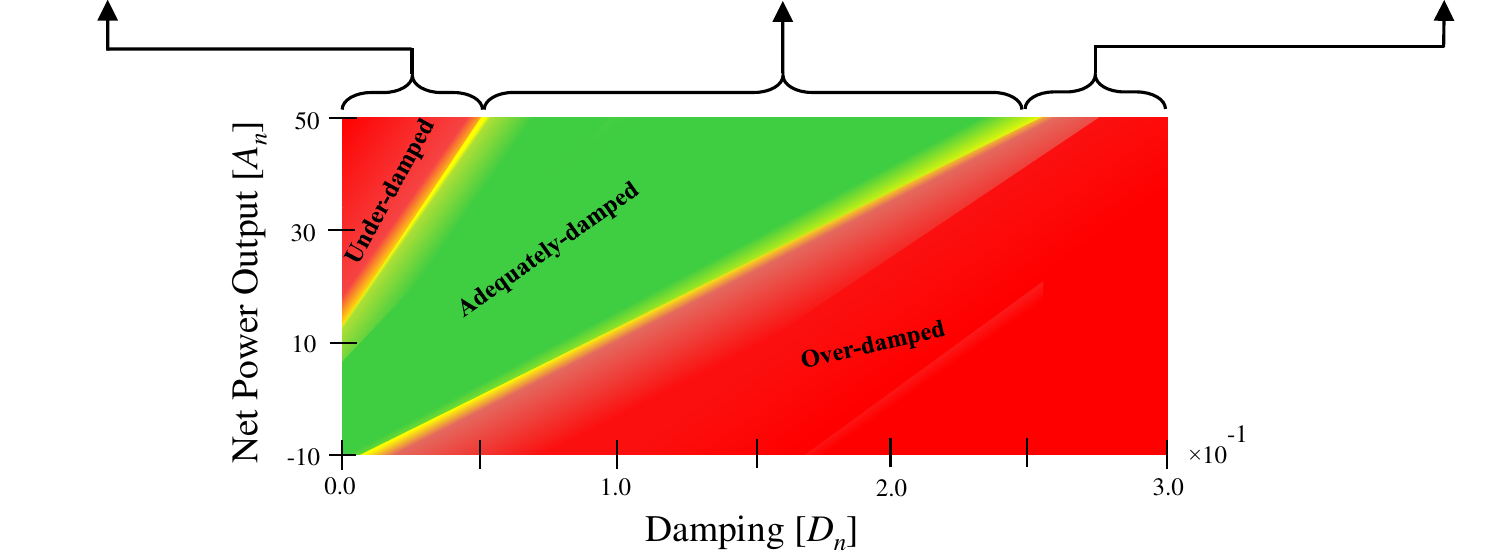} 
    \end{subfigure}  
    \caption{\textbf{The synchronization boundary for lunar power grid.} The central graphic in the $A_n$-$D_n$ plane shows the region of parameter space satisfying  Inequality (\ref{eq:unique_solution_precise_main}). The data in this figure was generated by simulating equations (1) of Supplementary Note 2, using $n=10$ generators, whose parameters are given in Supplementary Note 9. The three graphs show the result of numerical simulation of the swing equations (\ref{eq:Swing_Equation-simplified}) with $10$ generators. In each subfigure, the damping value of just a single generator was changed to give rise to an under-damped, adequately-damped, and over-damped scenario, respectively. Notice that both the under-damped and over-damped scenarios cause generator $10$ to enter into wild oscillations as a result of the infinite-period bifurcation. In the adequately damped case, notice that the solid state generators quickly synchronize with the central fission reactor, and the whole system then steadily increases towards the desired synchronization frequency of $1\text{kHz}\approx 6,283 \text{ rad}/\text{sec} $. For the given choice of parameters, the synchronization condition $|A_n - D_n \Delta\omega_{sync}|<K_{n1}$ yields the boundary lines $-12.7+247.08 D_n < A_n < 12.7+776.25 D_n$.
 We can see a single generator having damping outside of the allowable range can cause undesirable effects throughout the rest of the grid. In the over-damped case, the rogue generator's frequency is less than the rest of the generators, and in the under-damped scenario, the rogue generator's frequency is higher than the remaining generators. In both cases, because it continues to influence the rest of the grid, the rogue generator pulls the remaining generators out of equilibrium.
    }
     \label{fig:stability_diagram}
    \end{center}
\end{figure*}

Recall the notation $\mu_i=(D_i \Delta \omega_{sync}-A_i)/K_{1i}$ introduced in Eq. \eqref{eq:Fixed_Point_delta_i}. Examining equations (\ref{eq:(delta1)})-(\ref{eq:Fixed_Point_delta_i}) shows that as  $|\mu_i|$  goes from being less than $1$ to being greater than $1$ for some $i$, then the fixed points will pairwise collide and mutually annihilate, as can be seen by the fact that no fixed points exist when $|\mu_i|>1$. This suggests that a saddle node bifurcation  \cite{Strogatz_3rd_ed_2024} occurs at $|\mu_i|=1$. 
However, this is not the full picture. To fully understand the bifurcation, note that each $\delta_i$ is $2\pi$ periodic, in the sense that $\delta_i \equiv \delta_i +2n\pi$ for $n\in \mathbb{Z}$. For this reason, one can understand the dynamics of  $\delta_i$ as taking place on the circle $S^1,$ which can be understood as $[0,2\pi]$ with the endpoints glued together. Hence, the dynamics of $(\delta_1,\ldots,\delta_n)$ take place on the $n$-Torus $T^n=S^1\times\cdots \times S^1$.
After the saddle node bifurcation occurs, trajectories flow on the torus without ever converging to a fixed point. As it turns out, as illustrated in Figure \ref{fig:Bifurcation}, after the bifurcation, trajectories tend to a periodic motion on the torus. Thus, the bifurcation at $|\mu_i|=1$ is an \emph{infinite-period bifurcation}, in which a saddle-node bifurcation occurs on a periodic orbit. See \cite{Strogatz_3rd_ed_2024}, Chapter 8 for details on infinite period bifurcations.
The infinite-period bifurcation is a global bifurcation, in the sense that a small change in the network parameters leads to a large scale, global change in the dynamics. Instead of converging to a synchronized fixed point, after the bifurcation the system enters large amplitude periodic motions. These motions are pictured in Figure \ref{fig:stability_diagram}. This result explains why relatively small changes in the network parameters can lead to catastrophic failures in grid operations. Having identified the bifurcation which leads to the loss of grid stability, we next discuss implications of our results.


\section{Perspectives}\label{sec:Implications_Lunar}

An area of immediate impact for our results is the design of a passive control scheme to guarantee network stability. Compared to several recent proposed control schemes, which guarantee stable synchronization through active feedback control techniques \cite{Kasis_Passivity_Load_Participation, Trip_Burger_Passivity_Variable_Voltages,Schiffer_Port_Hamiltonian, Monshizadeh_Port_Hamiltonian_2, Monshizadeh_Port_Hamiltonian} our results provide a passive control scheme, in the sense that stability is guaranteed without the use of any active feedback monitoring. The simplicity of this control scheme is advantageous in the lunar power network setting, as it requires deploying fewer complex systems and instrumentation which are potential points of failure. This feature meets NASA's objective for future missions in Earth Independent Operations as it offers the ability to ``respond to on-board problems without the rapid response from a large number of ground support resources" \cite{doi:10.2514/6.2024-4869}. The design goal for mission controllers is to deploy loads and generation that respect condition \eqref{eq:unique_solution_precise_main} at all times. 

One notable implication of our results which we wish to point out is that stability of the grid is independent of the number $n$ of small generators. This is significant, as one of NASA's priorities for the lunar grid is that it be extensible, so that additional generators and assets can be added in a modular manner over time \cite{thomas2023modular}. Our work suggests that so long as the grid continues to develop in a hub-and-spokes topology, and so long as load deployment continuously respects Inequality (\ref{eq:unique_solution_precise_main}), components can be added without disrupting grid stability.

 Our results also predict a substantial challenge that will need to be overcome for the operation of the lunar grid. The coupling constants $K_{ij},$ which are so critical to the stability of the grid, tend to be inversely proportional to the operational frequency. (Recall that $K_{ij}$ depends linearly on the line admittance $Y_{ij},$ which itself is inversely proportional to the AC frequency.) Because of the necessity of operating at a frequency near 1 kHz, the coupling in the lunar grid will be decreased by a factor of roughly $16-20\times$ relative to the terrestrial grid, all else being equal. This fact will, according to Inequality (\ref{eq:unique_solution_precise_main}), tend to decrease the stability of the grid overall by the same order. This critical observation should be taken into consideration as NASA continues to study the specifications for the development of the lunar grid.


Another important implication of inequality (\ref{eq:unique_solution_precise_main}) which should be considered in designing the lunar power grid is that there are two ways in which a small generator can cause the grid to lose stability - either by being over-damped or by being under-damped. As illustrated in Figure \ref{fig:stability_diagram}, both scenarios can have equally dramatic effects on overall frequency stability, with both inducing large oscillations in the system frequency. On either side of stability the boundary, the ``rogue" generator decouples from the rest of the grid, entering large amplitude limit cycle oscillations. The boundaries of the adequately-damped region correspond to the infinite-period bifurcation described earlier.

Finally, note that Inequality (\ref{eq:unique_solution_precise_main}) can also be viewed as giving a bound on the maximum magnitude frequency deviation $\Delta \omega_{sync}$ that can occur in the grid while maintaining synchronization. Considering that the lunar grid will be operated with $\omega_R$ in the kHz range, the maximum relative frequency deviation $\Delta \omega_{synch}/\omega_R$ for which the system will continue to synchronize will be relatively small. This provides operational guidance for designing protection equipment in order to avoid losing synchronization and causing large oscillations.


The results here naturally can be extended to terrestrial power grids with solid state generators (known as inverter-based resources). Utilizing our method, we find that our main synchronization result is fundamentally independent of frequency of operation, and depends only on the grid topology. This observation enhances the applicability of our results, and verifies the efficacy of NASA's proposed design. By contrast with the lunar grid, the topology of modern terrestrial power grids is much more complex and dense. Therefore, it would be interesting to examine multiple connected hub-and-spoke systems together and  meshed networks, as these topologies more closely resemble potential long-term lunar power grid topology and that of the terrestrial grid.
Similarities and main differences in the principles of synchronization and operation the two grids are provided in Supplementary Notes 11 and 12.

We suggest that our paper has a broader impact beyond power networks across a wide range of disciplines as the methods developed here could also be applied to all complex networks that operate at high frequency and follow a hub-and-spoke model. 

\section{Acknowledgments}

This work was authored in part by the National Renewable Energy Laboratory, operated by Alliance for Sustainable Energy, LLC, for the U.S. Department of Energy (DOE). The views expressed in the article do not necessarily represent the views of the DOE or the U.S. Government. The U.S. Government retains and the publisher, by accepting the article for publication, acknowledges that the U.S. Government retains a nonexclusive, paid-up, irrevocable, worldwide license to publish or reproduce the published form of this work, or allow others to do so, for U.S.
This work was supported by the U.S. Department of Energy under Contract No. DE-AC36-08-GO28308 with the National Renewable Energy Laboratory.

Mark Walth acknowledges financial support by an appointment with the National Science Foundation (NSF) Mathematical Sciences
Graduate Internship (MSGI) Program. This program is administered by the Oak Ridge Institute for Science and Education (ORISE)
through an interagency agreement between the U.S. Department of Energy (DOE) and NSF. ORISE is managed for DOE by ORAU.

All opinions expressed in this paper are the author's and do not necessarily reflect the policies and views of NASA, NSF, ORAU/ORISE, or
DOE.

\bibliographystyle{IEEEtran}       
\bibliography{z_references}

\end{document}


\title{
Lunar Power Grid:\\
Network Structure and Spontaneous Synchronization\\ 
\vspace{1em}
\Large{Supplementary Information}
}

\vspace{1em}

\author{ 
M.~Walth, 
A.~Sajadi,
M.~Carbone,
and
B.~M.~Hodge

\thanks{M.~Walth is with the Mathematics Department at Cornell University, Ithaca, NY 14810, USA, Email: msw283@cornell.edu.}

\thanks{A.~Sajadi is with the Renewable and Sustainable Energy Institute (RASEI) at the University of Colorado Boulder, 4001 Discovery Drive, Boulder, CO 80303, USA, and the Power Systems Engineering Center (PSEC) at the National Renewable Energy Laboratory (NREL), 15013 Denver W Pkwy, Golden, CO 80401, USA, Email: Amir.Sajadi@colorado.edu, Amir.Sajadi@nrel.gov.}

\thanks{M.~Carbone is with the Power Management and Distribution Branch at the National Aeronautics and Space Administration (NASA) Glenn Research Center, 21000 Brookpark Rd, Cleveland, OH 44135, USA, Email: marc.a.carbone@nasa.gov.} 

\thanks{B.~M.~Hodge is with the Department of Electrical, Computer and Energy Engineering, the Department of Applied Mathematics, and the Renewable and Sustainable Energy Institute (RASEI) at the University of Colorado Boulder, 425 UCB, Boulder, CO 80309, USA, Email: BriMathias.Hodge@colorado.edu.}
}

\markboth{Preprint Draft, \today}%
{Sajadi \MakeLowercase{\textit{et al.}}: TBD}

\maketitle

\input{Nature_Physics_Sections/2_3_supp_lunar_grid}
\input{Nature_Physics_Sections/2_4_supp_dynamic_model}
\input{Nature_Physics_Sections/2_5_supp_necessary_condition}
\input{Nature_Physics_Sections/2_6_supp_linear_stability}
\input{Nature_Physics_Sections/2_7_supp_nonlinear_stability}
\input{Nature_Physics_Sections/2_9_supp_data}
\input{Nature_Physics_Sections/2_10_critical_damping}
\input{Nature_Physics_Sections/2_11_supp_implicatoins_earth}
\input{Nature_Physics_Sections/2_12_supp_differences_earth}

\bibliographystyle{IEEEtran}       
\bibliography{z_autosam}